\documentclass[AMA,STIX1COL,doublespace]{WileyNJD-v2}
\usepackage{geometry}
\usepackage{multirow}
\usepackage{amsmath}
\usepackage{amssymb}
\usepackage{verbatim}
\usepackage{graphicx}
\usepackage{float}
\usepackage{array}
\usepackage{caption}
\usepackage{subcaption,ulem,tikz}
\usepackage{rotating}
\usepackage{color}
\usepackage{pdflscape}
\usepackage{NJDnatbib}

\newcolumntype{H}{>{\setbox0=\hbox\bgroup}c<{\egroup}@{}}

\articletype{}%

\received{}
\revised{}
\accepted{}

\begin{document}

\title{Density Estimation in the Presence of Heteroscedastic Measurement Error of Unknown Type using Phase Function Deconvolution}

\author[1]{Linh Nghiem}

\author[1,2]{Cornelis J. Potgieter}

\authormark{Nghiem \& Potgieter (2017) }

\address[1]{\orgdiv{Department of Statistical Science}, \orgname{Southern Methodist University}, \orgaddress{\state{Texas}, \country{USA}}}

\address[2]{\orgdiv{Department of Statistics}, \orgname{University of Johannesburg} \orgaddress{ \country{South Africa}}}

\corres{Linh Nghiem. \email{lnghiem@smu.edu}}

\presentaddress{Heroy Hall 136B, 3225 Daniel Avenue, Dallas, TX 75206}

\abstract[Abstract]{It is important to properly correct for measurement error when estimating density functions associated with biomedical variables. These estimators that adjust for measurement error are broadly referred to as density deconvolution estimators. While most methods in the literature assume the distribution of the measurement error to be fully known, a recently proposed method based on the empirical phase function (EPF) can deal with the situation when the measurement error distribution is unknown. The EPF density estimator has only been considered in the context of additive and homoscedastic measurement error; however, the measurement error of many biomedical variables is heteroscedastic in nature. In this paper, we developed a phase function approach for density deconvolution when the measurement error has unknown distribution and is heteroscedastic. A weighted empirical phase function (WEPF) is proposed where the weights are used to adjust for heteroscedasticity of measurement error. The asymptotic properties of the WEPF estimator are evaluated. Simulation results show that the weighting can result in large decreases in mean integrated squared error (MISE) when estimating the phase function. The estimation of the weights from replicate observations is also discussed. Finally, the construction of a deconvolution density estimator using the WEPF is compared to an existing deconvolution estimator that adjusts for heteroscedasticity, but assumes the measurement error distribution to be fully known. The WEPF estimator proves to be competitive, especially when considering that it relies on minimal assumption of the distribution of measurement error.}


\maketitle
\newcommand{\Var}{\operatorname{Var}}
\newcommand{\Cov}{\operatorname{Cov}}

\doublespacing

\section{Introduction}
Many biomedical variables cannot be measured with great accuracy, leading to observations contaminated by measurement error.
Examples of such variables have been suggested in numerous epidemiological and clinical settings, including the measurement of blood pressure, radiation exposure, and dietary patterns. \cite{carroll2006measurement} The sources of measurement error range from the instruments used to measure the variables of interest to the inadequacy of short-term measurements for long-term variables; as such, the observed measurements have larger variance than the true underlying quantity of interest. The presence of measurement error can have a substantive impact on statistical inference. For example, not correcting for measurement error can result in biased parameter estimates, and loss of power in detecting relationships among variables. \cite{carroll2006measurement} Appropriate corrections need to be implemented when performing any data analysis with measurement error present to avoid making erroneous inferences.

A common problem of interest is to estimate the density of a variable when it is measured with additive measurement error. \cite{stirnemann2012density} This problem is often referred to as density deconvolution. When the noise-to-signal ratio is large, implementing a correction becomes crucial as the density of the observed data can deviate substantially from the true density of interest. Let $f_X(x)$ denote the density function of a random variable $X$, and assume that it is of interest to estimate $f_X(x)$ when $X$ is not directly observable. Specifically, we are only able to observe contaminated versions of $X$, say $W=X+U$, where $U$ represents measurement error. Thus, we are interested in estimating the density function of $X$ based on an observed sample $W_1, W_2, ..., W_n$ with $W_i = X_i + U_i,  i = 1, \ldots,n$. Here, the $X_i$ are an \textit{iid} sample from a distribution with density $f_X$, with $U_i$ representing the measurement error of the $i^{\mathrm{th}}$ observation. The $U_i$ are assumed both mutually independent and independent of the $X_i$. 

The nonparametric density deconvolution problem when first considered assumed that the distribution of the measurement error was fully known. \cite{carroll1988optimal}$^,$\cite{stefanski1990deconvolving} The development that followed in the literature mostly considered the case of known measurement error, and generally treated the measurement error as homoscedastic. \cite{fan1991asymptotic}$^,$\cite{fan1991optimal}$^,$ \cite{fan1993nonparametric}$^,$ \cite{hall2005discrete}$^,$ \cite{lee2010direct} The case of heteroscedastic measurement error was considered by Fan \cite{fan1992deconvolution} and Delaigle \& Meister. \cite{delaigle2008density} The problem of the measurement error having an unknown distribution was considered by Diggle \& Hall \cite{diggle1993fourier} and Neumann \&  H{\"o}ssjer, \cite{neumann1997effect} who assume that samples of error data are available, and by Delaigle et al.  \cite{delaigle2008deconvolution} who use replicate data to estimate the entire characteristic function of the measurement error. McIntyre \& Stefanski \cite{mcintyre2011density} considered the heteroscedastic case with replicate observations. Their work assumed the measurement errors all follow a normal distribution with unknown variances only. The phase function deconvolution approach developed by Delaigle \& Hall \cite{delaigle2016methodology} is groundbreaking in that they estimate the density function $f_X$ with both the measurement error distribution and variance unknown, and without the need for replicate data. Their method is based on minimal assumptions: The measurement error terms $U_i$ are only assumed to be mutually independent and  independent of the $X_i$ and to have a strictly positive characteristic function. However, Delaigle \& Hall \cite{delaigle2016methodology} only considered the case where the $U_i$ are homoscedastic, while heteroscedastic data is a reality often encountered in practice. In fact, the variance of measurement error often increases with the true underlying value.\cite{guo2011regression}

In this paper, we develop the phase function approach for density deconvolution when the measurement error has unknown distribution and is heteroscedastic. The model considered in this paper assumes the observed data are of the form $W_i = X_i + \sigma_i \varepsilon_i$ where the $X_i$ are an \textit{iid} sample from $f_X$, the measurement error terms $\varepsilon_i$ are independent and each $\varepsilon_i$ has a positive characteristic function and satisfies $\mathrm{E}(\varepsilon_i)=0$ and $\mathrm{Var}(\varepsilon_i)=1$. The $\sigma_i$ are non-negative constants and represent measurement error heteroscedasticity. Specifically, $\mathrm{Var}(W_i)=\sigma_X^2 + \sigma_i^2$ where $\sigma_X^2$ denotes the variance of $X$. Additionally, it is assumed that the random variable $X$ is asymmetric. This assumption is fundamental to the identifiability of the phase function of $X$, which forms the basis of estimation. A more detailed discussion of the model assumptions is presented in Section \ref{model_ass}, see also Delaigle \& Hall \cite{delaigle2016methodology}.

Note that the heteroscedasticity of the measurement error will require either that the constants $\sigma_i$ be known, or that there are replicate data so that the $\sigma_i$ can be estimated from the data. To illustrate the use of this estimator in a biomedical setting, a real-data example is included in Section 4. This example uses data from the Framingham Heart Study, which collected several variables related to coronary heart disease for study subset of $n=1615$ patients. For each patient, two measurements of long-term systolic blood pressure (SBP) were collected at each of two examinations. The distribution of true long-term SBP is estimated using the empirical phase function (EPF) and weighted empirical phase function (WEPF) density deconvolution estimator. These estimators are compared to a naive density estimator that makes no correction for measurement error, as well as the estimator of Delaigle \& Meister \cite{delaigle2008density} assuming the measurement error follows a Laplace distribution.

The remainder of the paper is organized as follows. Section 2 discusses the model assumptions,  considers estimation of the phase function and introduces a weighted empirical phase function (WEPF) which adjusts for heteroscedasticity in the data. A small simulation study compares two different weighting schemes. Section 3 shows how the WEPF can be inverted to estimate the density function $f_X$ and presents an approximation of the asymptotic mean integrated squared error for selecting the bandwidth. The WEPF deconvolution estimator is compared to that of Delaigle \& Meister, \cite{delaigle2008density} who treat the heteroscedastic case with known measurement error distribution. Section 4 illustrates the method using data from the Framingham Heart Study and Section 5 contains some concluding remarks.

\section{Phase Function Estimation}

\subsection{Model and Main Assumptions}\label{model_ass}

The model considered in the paper assumes the observed data are of the form $W_{i} = X_i + \sigma_i \varepsilon_{i}$ where the $X_i$ are an an \textit{iid} sample from $f_X$, the measurement error terms $\varepsilon_i$ are mutually independent and independent from $X_i$, and that each $\varepsilon_i$ has a strictly positive characteristic function. Note that the model does not require that the $\varepsilon_i$ have the same type of distribution, but only that each $\varepsilon_i$ has a characteristic function satisfying the above requirement. The assumption of a strictly positive characteristic function is equivalent to $\varepsilon_i$ being symmetric about zero with support on the entire real line. Many commonly used continuous distributions, including the Gaussian, Laplace, and Student's $t$ distributions, satisfy this assumption. In general, the only symmetric distributions excluded are those defined on bounded intervals (such as the uniform). For convenience, it is assumed that $\mathrm{Var}[\varepsilon_i]=1$, so that the constant $\sigma_i^2$ represents the heteroscedastic measurement error variance of the $i^{th}$ observation. Specifically, $\mathrm{Var}(W_i)=\sigma_X^2 + \sigma_i^2$ where $\sigma_X^2$ denotes the variance of $X$. The density function $f_X$ is assumed to be asymmetric. More specifically, it is assumed that the random variable $X$ does not have a symmetric component. This means that there is \textit{no} symmetric random variable $S$ for which $X$ can be decomposed as $X=X_0+S$ for arbitrary random variable $X_0$. This asymmetry is crucial to the ability to estimate the true density function of $X$. As discussed in Delaigle \& Hall,\cite{delaigle2016methodology} if one were to assumed that the density function $f_X$ were sampled from a random universe of distributions, then the assumption of indecomposability is satisfied with probability $1$. Practically, the indecomposability assumption is not unreasonable as data are rarely observed from a perfectly symmetric distribution. There is a special type of distribution for $X$ that cannot be recovered by this method, namely when $X$ is itself a convolution (sum) of a skew distribution and a symmetric distribution. The result from Delaigle \& Hall indicates that this need not be a concern for the general practitioner implementing this method.  While the exposition in this paper assumes that the measurement error components are independent, the methodology could be generalized to a setting where $\mathrm{Cov}[\varepsilon_i,\varepsilon_j]=\sigma_{ij}\neq 0$ for some pairs $i\neq j$. This would not affect the proposed estimator directly, but would have consequences for how the bandwidth is chosen. The latter question is beyond the scope of the present paper.

\subsection{The Weighted Empirical Phase Function (WEPF)}

The phase function of a random variable $X$, denoted $\rho_X(t)$, is defined as the characteristic function of $X$ standardized by its norm,
\begin{equation}
\rho_X (t) = \dfrac{\phi_X(t)}{\vert\phi_X(t)\vert} \label{phase_func_def}
\end{equation}
with $ \phi_Z(t)$ the characteristic function of a random variable $Z$ and $\vert z \vert = (z \bar{z})^{1/2}$ denoting the norm function with $\bar{z}$ the complex conjugate of $z$. Let $W=X+\sigma \varepsilon$ with $\varepsilon$ having characteristic function $\phi_{\varepsilon}(t) \ge 0$ for all $t$. It is easy to verify that the random variables $W$ and $X$ have the same phase function, $\rho_W(t) = \rho_X(t)$. Delaigle \& Hall \cite{delaigle2016methodology} use this relation and an empirical estimate of $ \phi_W(t)$ in equation (\ref{phase_func_def}) to estimate the phase function, see their paper for details on implementation.  

In the case of heteroscedastic errors, we propose to use a weighted empirical phase function (WEPF) to adjust for heteroscedasticity. Define function
\begin{equation}
\hat{\phi}_W(t|\boldsymbol{q}) = \sum_{j=1}^n q_j \exp (itW_j) \label{pseudo-cf}
\end{equation}
where $\boldsymbol{q}=\{q_1,\ldots,q_n\}$ denotes a set of non-negative constants that sum to $1$. Function (\ref{pseudo-cf}) is a weighted empirical characteristic function and noting random variable $W_i = X_i + \sigma_i \varepsilon_i$ has characteristic function $\phi_{W_i}(t) = \phi_X(t) \phi_{\varepsilon_i}(\sigma_i t)$, $i = 1,\ldots,n$, it follows that
\[\mathrm{E}[\hat{\phi}_W(t|\boldsymbol{q})] = \phi_X(t) \sum_{j=1}^n q_j \phi_{\varepsilon_{j}} (\sigma_j t). \]
The WEPF is defined as
\begin{equation}
\hat{\rho}_W(t|\boldsymbol{q}) = \frac{\hat{\phi}_W(t|\boldsymbol{q})}{\vert \hat{\phi}_W(t|\boldsymbol{q}) \vert} = \frac{\sum_{j} q_j \exp (itW_j)}{\left\{\sum_{j} \sum_{k} q_j q_k \exp [it(W_j-W_k)]\right\}^{1/2}} .
\label{eq:WEPF}
\end{equation}  
For $\boldsymbol{q}_{eq}=\{1/n,\ldots,1/n\}$, $\hat{\rho}_W(t|\boldsymbol{q}_{eq})$ essentially reduces to the phase function proposed by Delaigle \& Hall.\cite{delaigle2016methodology} Use of weights choice $\boldsymbol{q}_{eq}$ will be referred to as the empirical phase function (EPF) estimator. Other choices of weights can serve as an adjustment for heteroscedasticity -- observations with large measurement error variance can be down-weighted to have smaller contribution to the phase function estimate.

The asymptotic properties of the WEPF are given in the Theorem \ref{theorem:WEPF} below.
\begin{theorem}
	Assume that $\max_j q_j = \mathcal{O}(n^{-1})$ and that each measurement error component $\varepsilon_{j}$ has a strictly positive characteristic function. It then follows that the WEPF as defined in \eqref{eq:WEPF} is a consistent estimator of the phase function of $W$, and hence of the phase function of $X$. Also, the asymptotic variance of the WEPF is given by	
	\begin{align}
	\mathrm{AVar}[\hat{\rho}_W\left(t\vert \boldsymbol{q}\right)-\rho_W\left(t\right)] 
	&= \frac{1}{2\left\vert \phi _{X}\left( t\right) \right\vert ^{2}\psi
		_{\varepsilon }\left(t\vert \boldsymbol{q}\right) }\sum_{k=1}^{n}q_{k}^{2}\left[ 1-\left\vert
	\phi _{X}\left( t\right) \right\vert ^{2}\phi _{\varepsilon _{k}}^{2}\left(
	\sigma _{k}t\right) +\phi _{\varepsilon _{k}}^{2}(\sigma _{k}t)\right] \nonumber \\ 
	&\quad -\frac{\mathrm{Re}\left\{ \phi _{X}^{2}\left( t\right) \phi _{X}\left(
		-2t\right) \right\} }{2\left\vert \phi _{X}\left( t\right) \right\vert
		^{4}\psi _{\varepsilon }\left(t\vert \boldsymbol{q}\right) }\sum_{k=1}^{n}q_{k}^{2}\phi
	_{\varepsilon _{k}}(2\sigma _{k}t)
	\label{eq:var1}
	\end{align}
	where  $ \psi _{\varepsilon }\left(t\vert \boldsymbol{q}\right) = \left[\sum_{k}q_{k}\phi _{\varepsilon _{k}}(\sigma
	_{k}t)\right]^2.$
	\label{theorem:WEPF}
\end{theorem}

The proof of Theorem \ref{theorem:WEPF} can be found in the Supplementary Material. Equation \eqref{eq:var1} shows that the asymptotic variance of $\hat{\rho}_W(t|\boldsymbol{q})$ depends on $\phi_{\varepsilon_j}(t)$ $j=1,\ldots,n$, the characteristic functions of the measurement error components. While one would ideally like to choose weights $\boldsymbol{q}$ that minimize said asymptotic variance, this is unrealistic as the method proposed in this paper makes no parametric assumptions about the measurement error, meaning the $\phi_{\varepsilon_{j}}$ are unknown. A much simpler weighting scheme is proposed here, relying only on knowledge of the measurement error variances.

Note that $\mathrm{E}(W_i) = \mathrm{E}(X) = \mu$. As such, for weights $\boldsymbol{q}$, the estimator $\hat{\mu}_{\boldsymbol{q}} = \sum_{j=1}^{n} q_j W_j$ is an unbiased estimator of $\mu$. The weights
\begin{equation}
q_{i}^* = {\sigma^{-2}_{W_i}} \Big[\sum_{j=1}^{n} {\sigma^{-2}_{W_j}}\Big]^{-1} = (\sigma_X^2+\sigma_i^2)^{-1}\Big[\sum_{j=1}^{n} {(\sigma_X^2+\sigma_j^2)^{-1}}\Big]^{-1}
\label{eq:mean-optimal}
\end{equation}
result in a minimum variance estimator of $\mu$. This does have a connection to the phase function, as $\rho'_X(0) = \mu$; see the supplemental material of Delaigle \& Hall \cite{delaigle2016methodology} for the connection between the phase function and the odd moments of the underlying distribution. Let $\boldsymbol{q}_{opt}=\{q_1^*,\ldots,q_n^*\}$ denote the vector of mean-optimal weights and let WEPF$_{opt}$ denote the weighted empirical phase function estimator calculated using the mean-optimal weights. Both the performance of the EPF and the WEPF$_{opt}$ will be considered for estimating the phase function and density function.

\subsection{Estimating the Variance Components} \label{Estimating Variances}

In practice, it is often the case that neither the measurement error variances $\sigma_1^2,\ldots,\sigma_n^2$ nor $\sigma_X^2$ is known. These quantities can be easily estimated from replicate observations. This section describes how to estimate the variance components for a heteroscedastic measurement error variance model. In a setting where the underlying measurement error variance structure is unknown, the procedure outlined in this section can be used to estimate the mean-optimal weights in \eqref{eq:mean-optimal} used for estimating the WEPF.

Consider replicate observations, $
W_{ij}=X_{i}+\tau_{i}e _{ij}$, $j=1,\ldots,n_i$, $i=1,\ldots,n$ with $\min_i n_i \ge 2$, $E(e_{ij})=0$, $\mathrm{Var}(e_{ij})=1$, and $\tau_i^2$ representing heteroscedastic measurement error variance at the observation level. Note that $W_{ij}-W_{ij^{\prime }}=\tau_{i}\left( e _{ij}-e_{ij'}\right)$ and thus $\mathrm{E}\left[ \left( W_{ij}-W_{ij^{\prime }}\right) ^{2}\right] =2\tau_{i}^{2}$ for $j\neq j'$.
Define grand mean
\begin{equation*}
\bar{W}	= \dfrac{1}{n} \sum_{i=1}^{n}\left[ \dfrac{1}{n_i} \sum_{j=1}^{n_i} W_{ij} \right] = \dfrac{1}{n} \sum_{i=1}^{n}{X_i}+\dfrac{1}{n} \sum_{i=1}^{n} \left[\dfrac{\tau_i}{n_i} \sum_{j=1}^{n_i}  e_{ij}\right]
\end{equation*}
and note that $\mathrm{E}(\bar{W}) = \mu$
and
\[
\Var(\bar{W}) = \dfrac{\sigma_X^2}{n} + \dfrac{1}{n^2}\sum_{i=1}^{n}\dfrac{\tau_i^2}{n_i}.
\]
It can also be shown that
\begin{equation}
E\left[ \left( W_{ij}-\bar{W}\right) ^{2}\right]  = \sigma _{X}^{2}+\tau_{i}^{2} + \mathcal{O}(n^{-1}). \label{variance_eq}
\end{equation}
Subsequently, the variance components can be estimated by
\[
\hat{\tau}_{i}^{2}=\frac{1}{n_i\left( n_i-1\right) }\sum_{j=1}^{n_i-1}\sum_{j^{%
		\prime }=j+1}^{n_i}\left( W_{ij}-W_{ij^{\prime }}\right) ^{2}, \quad i=1,\ldots,n,
\]
and, motivated by \eqref{variance_eq},
\[\hat{\sigma}_{X}^{2} = \dfrac{1}{N} \sum_{i=1}^{n} \sum_{j=1}^{n_i} (W_{ij}-\bar{W})^2 - \dfrac{1}{n} \sum_{i=1}^{n} \hat{\tau}_i^2
\]
with $N=\sum_i n_i$. The analysis then proceeds by defining individual-level averages $W_{i}=(n_i^{-1})\sum_{j=1}^{n_i}W_{ij}$ and noting that $W_i = X_i + \sigma_i \varepsilon_i$ where $\sigma_i = \tau_i/\sqrt{n_i}$ and $\varepsilon_i$ has a distribution with a positive characteristic function whenever the same is true for all elements of the set $\{e_{i1},\ldots,e_{in_i}\}$. The estimate of $\sigma_i$ is given by $\hat{\sigma}_i = \hat{\tau}_i / \sqrt{n_i}$.

\subsection{Simulation Study} \label{Phase_Func_Sim}

A small simulation study was conducted to compare the performance of the EPF and WEPF$_{opt}$ estimators. The true $X_i$ data were sampled from the following three distributions: (1) $X \sim \chi^2_3/\sqrt{6}$ (Scaled $\chi^2_3$), (2) $X \sim \left(0.5 \text{N}(1,1) + 0.5 \chi^2 (5) \right)/\sqrt{9.5}$ (Mixture 1), and (3) $X \sim \left(0.5 \text{N}(5,0.6^2) + 0.5 \text{N}(2.5,1)\right)/\sqrt{2.2425} $ (Mixture 2). The first two distributions are right-skewed while the third distribution is bimodal. All three distributions were scaled to have unit variance. The phase functions of these distributions are shown in Figure 1 of the Supplemental Material. The measurement error terms $\varepsilon_{ij}=\tau_i e_{ij}$ were sampled from a normal distribution with mean 0 and variance structure $\tau_{i}^2 = J\sigma_i^2$ with $ \sigma_i^2 = 0.025 \sigma^2_X, i=1,\ldots, n/2$ and $\sigma_i^2 = 0.975 \sigma^2_X, i= n/2+1,\ldots, n$. For each candidate distribution of $X$, a total of $N=1000$ samples $W_{ij}=X_i + \tau_ie_{ij}$, $i=1,\ldots,n$ and $j=1,\ldots,J$ were generated for sample sizes  $n = 250, 500, \text{ and } 1000$. Scenarios with no replicates ($J=1$) and also with replicates ($J=2$ and $3$) were considered in the simulation. Under the scenario with no replication, the measurement error variance was treated as known. In settings with $J=2$ and $3$ replicates, the measurement error variances were estimated from the replicate data using the procedure outlined in Section \ref{Estimating Variances}. The choice of observation-level measurement error variance $\tau_i^2 = J\sigma_i^2$ results in the combined replicate values $W_i = J^{-1} \sum_j W_{ij}$ having measurement error variance $\sigma_i^2$. This was done to make the simulation results with and without replicates easily comparable. For each simulated dataset, the mean-optimal weight vector $\boldsymbol{q}_{opt}$ was calculated (or estimated in the case of replicate data) using equation \eqref{eq:mean-optimal}. The WEPF$_{opt}$ estimator was then calculated using these weights. Additionally, the EPF estimator was calculated using equal weights for all observations. As the quality of the empirical characteristic function decreases with increasing $t$, the suggestion of  Delaigle \& Hall \cite{delaigle2016methodology} was followed and the estimated phase functions were only computed on the interval [$-t^*, t^*$], where $t^*$ is the smallest $t>0$ such that $|\hat{\phi}_W(t|\boldsymbol{q})| < n^{-1/4}$. The EPF and WEPF are compared using (estimated) mean integrated squared error (MISE) ratios, $\mathrm{MISE}_{eq}/\mathrm{MISE}_{opt}$, where $\mathrm{MISE}_{eq}$ and $\mathrm{MISE}_{opt}$ denote the MISEs of the EPF and WEPF$_{opt}$ estimators respectively. The results are summarized in Table \ref{tab:MISE phase}.

\begin{table}[h]
	\centering	
	\begin{tabular}[h]{|c|l|c|c|c|}
		\hline
		Replicates & Distribution & $n=250$ & $n=500$ & $n=1000$  \\
		\hline
		No replicate &$X \sim \chi^2_3/\sqrt{6}$ &  1.220 (0.021)  & 1.280 (0.020)  & 1.277 (0.023) \\
		& $X \sim $ Mixture 1 &  1.298 (0.023) & 1.321 (0.022) &  1.303 (0.022) \\
		&$X \sim $ Mixture 2
		&  1.065 (0.017) &  1.085 (0.018) & 1.109 (0.019) \\
		\hline
		2 replicates &$X \sim \chi^2_3/\sqrt{6}$ &  1.075 (0.016)  & 1.155 (0.018)  & 1.139 (0.018) \\
		& $X \sim $ Mixture 1 &  1.044 (0.007) & 1.021 (0.006) &  1.005 (0.004) \\
		&$X \sim $ Mixture 2
		&  1.003 (0.004) & 1.007 (0.003) & 1.007 (0.002) \\
		\hline
		3 replicates &$X \sim \chi^2_3/\sqrt{6}$ &  1.150 (0.019)  & 1.177 (0.019)  & 1.150 (0.020) \\
		& $X \sim $ Mixture 1 &  1.020 (0.008) & 1.017 (0.006) &  1.001 (0.004) \\
		&$X \sim $ Mixture 2
		& 1.001 (0.004) & 1.005 (0.003) & 1.008 (0.002) \\
		\hline
		
	\end{tabular}  
	\caption{The ratio $\mathrm{MSE}_{eq} / \mathrm{MSE}_{opt}$ and the corresponding jackknife standard error (in parentheses) when estimating the phase function of $X$ with normal measurement error and variance structure given in Case 1 of Table \ref{table:error}, based on $N=1000$ samples, when there are no replicate (assuming the true variances of measurement errors are known), 2 replicates, and 3 replicates per observation.}
	\label{tab:MISE phase}
\end{table} 

In Table \ref{tab:MISE phase}, an MISE ratio greater than $1$ indicates better performance of the WEPF$_{opt}$ estimator compared to the EPF estimator. The table also reports estimated standard errors for the MISE ratios. The standard errors were estimated using the following jackknife procedure. For the $j^{th}$ simulated sample, let $(\mathrm{ISE}_{eq,j},\mathrm{ISE}_{opt,j})$ denote the integrated squared error for the EPF and the WEPF$_{opt}$ respectively,  $j = 1,\ldots,N$. Let $R_{(-j)}$ denote the MISE ratio calculated after deleting the $j^{th}$ ISE pair. Then, the jackknife standard error for the MISE ratio is given by
\[
\mathrm{SE}_{jack} = \sqrt{\dfrac{1}{N}{\sum_{j=1}^{N}}\left(R_{(-j)}-\bar{R} \right)^2}
\]
where $\bar{R} = N^{-1} \sum_{j=1}^{N} R_{(-j)} $.

Inspection of Table \ref{tab:MISE phase} shows that the WEPF$_{opt}$ performs better than the EPF for the measurement error configuration considered. When the measurement error variances are known, the gain from using WEPF$_{opt}$ can be substantial. Specifically, the MISE of WEPF$_{opt}$ is seen to between $6.5\%$ and $30\%$ lower than the MISE of the EPF for the distributions considered. When there are $J=2$ and $J=3$ replicates per observation, the WEPF$_{opt}$ performs slightly better than the EPF for the scaled $\chi^2_3$ distribution, while their performance is nearly identical for Mixtures 1 and 2. In this setting, the use of the suggested weighting scheme never results in poorer performance of the WEPF$_{opt}$ estimator compared to the EPF estimator.

Next, the effect of different underlying measurement error variance structures on the MISE ratio of the EPF and WEPF$_{opt}$ was examined. The sample size was fixed at $n=1000$ and the three different measurement error variance structures considered are outlined in Table \ref{table:error}. The ratios $\mathrm{MSE}_{eq} / \mathrm{MSE}_{opt}$ based on $1000$ simulated datasets are reported in Table \ref{MISE phase_short}. Again, jackknife estimates of standard error are also reported. 

\setlength\extrarowheight{5pt}
\begin{table}[h]
	\centering
	\begin{tabular}{|p{2cm}|p{12cm}|}
		\hline
		Case & Variance Structure \\
		\hline 
		Case 1 & $ \sigma_i^2 = 0.025 \sigma^2_X, i=1,\ldots, n/2$ and $\sigma_i^2 = 0.975 \sigma^2_X, i= n/2+1,\ldots, n$ \\
		Case 2 & $\sigma_i^2 = (0.25+ 0.5i/n) \sigma_X^2$, $i = 1,\ldots, n$\\
		Case 3 & $\sigma_i^2 = (0.025+ 0.95i/n) \sigma_X^2$, $i = 1,\ldots, n$ \\
		\hline 	
	\end{tabular}
	\caption{Three measurement error variance structures used in simulations.}
	\label{table:error} 
\end{table}
\setlength\extrarowheight{0pt}

\begin{table}[h]
	\centering	
	\begin{tabular}{|c|c|c|c|c|}
		\hline
		Replicates &$X$ & Case 1 & Case 2 & Case 3 \\
		\hline
		No replicate &$X \sim \chi^2_3 / \sqrt{(6)}$ & 1.277 (0.023) &  1.030 (0.005) & 1.113 (0.002) \\
		&$X \sim $ Mixture 1 &  1.303 (0.022) & 1.027 (0.006)  & 1.117 (0.012) \\
		&$X \sim $ Mixture 2 & 1.109 (0.019) & 1.011 (0.006) & 1.039 (0.012) \\
		\hline
		2 replicates &$X \sim \chi^2_3 / \sqrt{(6)}$ & 1.139 (0.018) &  0.925 (0.014) & 0.978 (0.015) \\
		&$X \sim $ Mixture 1 &  1.005 (0.004) & 0.992 (0.005)  & 0.998 (0.004) \\
		&$X \sim $ Mixture 2 & 1.007 (0.002) & 1.001 (0.003) & 1.002 (0.002) \\
		\hline
		3 replicates &$X \sim \chi^2_3 / \sqrt{(6)}$ & 1.150 (0.020) &  0.965 (0.014) & 1.034 (0.016) \\
		&$X \sim $ Mixture 1 &  1.001 (0.004) & 0.994 (0.004)  & 0.998 (0.004) \\
		&$X \sim $ Mixture 2 & 1.008 (0.002) & 0.999 (0.002) & 1.002 (0.002) \\
		
		\hline
	\end{tabular}  
	\caption{The effect of the error variance structure on the ratio $\mathrm{MISE}_{eq} / \mathrm{MISE}_{opt}$ and the corresponding jackknife standard error (in parentheses) based on $1000$ samples of size $n=1000$.}
	\label{MISE phase_short}
\end{table}

Inspection of Table \ref{MISE phase_short} illustrates the effect of different heterogeneity patterns of measurement error variances on the performance of the EPF and WEPF$_{opt}$ estimators. When the measurement error variances are known ($J=1$), the WEPF$_{opt}$ has a lower MISE than the EPF in all the considered configurations, with the heterogeneity pattern only affecting the size of the improvement. In the case of $J=2$ replicates per observation, there were four instances in Case 2 and Case 3 of measurement error variances where the EPF performed better than the WEPF$_{opt}$. This occurrence was likely because the estimated weights for WEPF$_{opt}$ were calculated from estimated variance components based on only a small number of replicates. When the number of replicates increases from $J=2$ to $J=3$, measurement error variances are estimated with higher accuracy, so the MISE ratio increase in general. Note that, although using WEPF$_{opt}$ can sometimes lead to a worse performance, the loss tends to be small (at most 8\% as seen in the Case 2 measurement error variance setting when $X$ follows a Scaled-$\chi_3^2$ with $2$ replicates); however, using WEPF$_{opt}$ can still result in large gains (as much as 15\% in the Case 1 measurement error variance setting when $X$ follows a Scaled-$\chi_3^2$ with $3$ replicates).

In general, the simulation study shows that weighting to adjust for heteroscedasticity in estimating the phase function never results in a much poorer estimator, but sometimes leads to a large gain in efficiency. The loss/gain depends on how accurate measurement error variances were estimated as evidenced by the improvement in going from $J=2$ to $J=3$ replicates. In the next section, this is explored in the context of density deconvolution. 

\section{Density Estimation}

\subsection{Constructing an Estimator of $f_X$} \label{method_decon}
The outline here is a brief overview of how the method of  Delaigle \& Hall \cite{delaigle2016methodology} can be implemented using the WEPF to estimate the density function $f_X$. Let $\hat{\phi}_W(t|\boldsymbol{q})$ and $\hat{\rho}_W(t|\boldsymbol{q})$ denote the weighted empirical characteristic function and corresponding WEPF respectively. Let $w(t)$ denote a non-negative weight function. Also let $x_j$, $j=1,\ldots,m$ denote a set of arbitrary values with respective probability masses $p_j$.  Delaigle \& Hall suggest a two-stage estimation method for $f_X$. First, one finds a characteristic function of the form $\psi(t|\mathbf{x,p})=\sum_j p_j \exp (itx_j)$ that has phase function close to the WEPF. Since this characteristic function corresponds to a discrete distribution with probability mass $p_j$ at the point $x_j$ for $j=1,\ldots,m$, the second stage of estimation involves smoothing $\psi(t|\mathbf{x,p})$ before applying an inverse Fourier transformation to obtain the estimated density $\hat{f}_X(x)$.  Delaigle \& Hall suggest sampling the $x_j$ uniformly on the interval [min $W_i$, max $W_i$] with $m = 5\sqrt{n}$. The goal is then to find the set $\{p_j\}_{j=1}^{m}$ that minimizes 
\begin{equation}
T(\boldsymbol{p}) = \int_{-\infty}^{\infty} \left|\hat{\rho}_W(t|\boldsymbol{q}) - \dfrac{\psi(t|\mathbf{x,p})}{\vert\psi(t|\mathbf{x,p})\vert}\right|^2 w(t) dt \label{eq:T_p}
\end{equation}
under the constraint of also minimizing the variance of the corresponding discrete distribution, 
$v(\mathbf{p}) = \sum_{j=1}^{m} p_j x_j^2 - (\sum_{j=1}^{m} p_j x_j)^2 $. This non-convex optimization problem of finding the solution $\{\hat{p}_j\}_{j=1}^{m}$ can be solved using MATLAB. Details are given in Delaigle \& Hall. \cite{delaigle2016methodology} The present implementation differs only in that the estimated phase function is weighted to adjust for heteroscedasticity. Beyond using a different estimator of the phase function, the optimization problem remains unchanged.

Now, let $\psi(t|\mathbf{x},\hat{\boldsymbol{p}}) = \sum_j \hat{p}_j \exp(itx_j)$ be the characteristic function with the $\hat{p}_j$s the probability masses estimated by minimizing \eqref{eq:T_p}. The deconvolution density estimator based on the WEPF is then
\begin{equation}
\hat{f}_{X}\left( x\right) =\frac{1}{2\pi }\int \exp \left( -itx\right) 
\tilde{\phi}\left( t\right) K^{\text{ft}}\left( ht\right) dt
\label{eq:dens}
\end{equation} 
where 
\[
\tilde{\phi}(t) =
\begin{cases}
\psi(t|\mathbf{x},\hat{\boldsymbol{p}}), & \text{for } t \leq t^* \\
r(t), & \text{for } t > t^*
\end{cases} 
\]
with $t^*$ being the smallest $t>0$ such that $|\hat{\phi}_W(t|\boldsymbol{q})| < n^{-1/4}$. Here, $K^{\text{ft}}(t)$ denotes the Fourier transform of a deconvolution kernel function and $r(t)$ denotes a ridging function. The ridging function ensures that the estimator is well-behaved outside the range $[-t^*,t^*]$. The proposed choice of ridging function is  $r(t) = \hat{\phi}_W(t|\boldsymbol{q})/\hat{\phi}_{L}(t)$, with $\hat{\phi}_{L}(t)$ the characteristic function of a Laplace distribution with variance equal to an estimator of $\sigma_L^2=\sum_{j} q_j \sigma_j^2$, the weighted sum of the measurement error variances. In application here, the common choice $K^{\text{ft}}(t) = (1-t^2)^3$ for $|t|\leq 1$ is used. The weight function is chosen to be $w(t)=\omega(t)\vert \hat{\phi}_W(t|\mathbf{q})\psi(t|\mathbf{x,p})\vert^2$ with $\omega(t)=0.75(1-t^2)$ for $|t|\leq 1$ (the Epanechnikov kernel) rescaled to the interval $[-t^*,t^*]$. This choice of weight function avoids numerical difficulties that can arise when dividing by very small numbers.

\subsection{Bandwidth Selection}
The proposed phase function deconvolution estimator that accounts for heteroscedasticity in \eqref{eq:dens} is an approximation of the estimator
\begin{equation}
\tilde{f}\left( x\right) =\frac{1}{2\pi }\int \exp{(-itx)}K^{\text{ft}}\left( ht%
\right) \frac{\hat{\phi}_W(t|\boldsymbol{q}) }{\sum_j q_{j}\phi _{\varepsilon
		_{j}}( \sigma _{j}t) }dt \label{approx estim}
\end{equation}
with $\hat{\phi}_W(t|\boldsymbol{q})$ defined in (\ref{pseudo-cf}). Note that (\ref{approx estim}) is an estimator that one could compute if the measurement error distribution were known, but that it is different from the heteroscedastic estimator proposed by  Delaigle \& Meister. \cite{delaigle2008density}
Taking expectation of the integrated squared error (ISE) of (\ref{approx estim}), $\mathrm{ISE} =\int [ \tilde{f}\left( x\right) -f_{X}\left( x\right)
] ^{2}dx$, gives mean integrated squared error (MISE)
\begin{eqnarray}
\mathrm{MISE} &=&\frac{1}{2\pi }\int \left\vert \phi _{X}\left( t\right) \right\vert ^{2}%
\left[ K^{\text{ft}}\left( ht\right) -1\right] ^{2}dt+\frac{1}{2\pi }\int
\left[ K^{\text{ft}}\left( ht\right) \right]^2 \frac{\sum_{j} q_j^2}{%
	\big[\sum_j q_{j}\phi _{\varepsilon _{j}}\left(\sigma _{j}t\right)\big]
	^{2}}dt \notag \\
&&-\frac{1}{2\pi }\int \left\vert \phi _{X}\left( t\right) \right\vert
^{2}\left[ K^{\text{ft}}\left( ht\right) \right]^2 \frac{ \sum_{j}q_{j}^{2}\phi
	_{\varepsilon _{j}}^{2}\left( \sigma _{j}t\right) }{\left[ \sum_j
	q_{j}\phi _{\varepsilon _{j}}\left( \sigma _{j}t\right) \right] ^{2}}dt. \label{eq:MISE}
\end{eqnarray}%
An argument similar to that of  Delaigle \& Meister \cite{delaigle2008density} when evaluating the asymptotic MISE (AMISE) of their heteroscedastic estimator, one can show that the last term of \eqref{eq:MISE} is negligible, giving
\[
\mathrm{AMISE}=\frac{1}{2\pi }\int \left\vert \phi _{X}\left( t\right)
\right\vert ^{2} \left[ K^{\text{ft}}(ht) -1\right] ^{2}dt+\frac{1}{%
	2\pi }\int \left[K^{\text{ft}}(ht)\right]^2 \frac{ \sum_{j}q_{j}^{2}%
}{\left[ \sum_j q_{j}\phi _{\varepsilon _{j}}\left( \sigma
_{j}t\right) \right] ^{2}}dt
\]%
In the present application, both $\phi _{X}\left( t\right) $ and $\phi
_{\varepsilon _{j}}\left( t\right) $, $j=1,\ldots,n$ are unknown. However,
note that $\left\vert \phi _{X}\left( t\right) \right\vert ^{2}=\phi
_{X}\left( t\right) \phi _{X}\left( -t\right) $ is the characteristic
function of the random variable $X-X^{\prime }$, where $X$, $X^{\prime }$
are \textit{iid} $f_{X}$. Regardless of the shape of $f_{X}$, the
random variable $X-X^{\prime }$ is symmetric about 0 and has variance $%
2\sigma _{X}^{2}$. This suggests replacing $\left\vert \phi _{X}\left(
t\right) \right\vert ^{2}$ with the characteristic function of a symmetric
distribution with mean $0$ and variance $2\hat{\sigma}_{X}^{2}$. Appropriate
choices might be the normal distribution, i.e. substituting $\exp \left( -%
\hat{\sigma}_{X}^{2}t^{2}\right) $ for $\left\vert \phi _{X}\left( t\right)
\right\vert ^{2}$, or the Laplace distribution, i.e. substituting $\left( 1+%
\hat{\sigma}_{X}^{2}t^{2}\right) ^{-1}$. Additionally, one can use
appropriate approximations for $\phi _{\varepsilon _{j}}\left( \sigma
_{j}t\right) $. For example, the Laplace choice is a reasonable one.
\cite{meister2006density}\cite{delaigle2008alternative} One can therefore substitute $\left(
1+0.5\hat{\sigma}_{j}^{2}t^{2}\right) ^{-1}$ for $\phi _{\varepsilon
	_{j}}\left( \sigma _{j}t\right) $. 
This Normal-Laplace substitution gives approximate AMISE function%
\begin{align}
\mathrm{\hat{A}}\left( h\right) =& \quad \frac{1}{2\pi }\int \exp \left( -\hat{\sigma%
}_{X}^{2}t^{2}\right) \left[ K^{\text{ft}}\left( ht\right) -1\right] ^{2}dt \nonumber \\
& +\frac{1}{2\pi }\int \left[ K^{\text{ft}}\left( ht\right) \right]^2 \frac{
	\sum_{j}q_{j}^{2} }{\left[ \sum_j q_{j}\left( 1+0.5\hat{\sigma}%
	_{j}^{2}t^{2}\right) ^{-1}\right] ^{2}}dt \label{bandwidth_calc}
\end{align}
and the value of $h$ that minimizes the above function can then be used to evaluate the density deconvolution estimator in equation \eqref{eq:dens}.%

\subsection{Simulation Study} \label{Sim_Dens}
Simulation studies were done to evaluate the performance of the equally-weighted and mean-optimal weighted phase function deconvolution density estimators. These correspond to the use of the EPF and WEPF$_{opt}$ as the phase function estimate before performing the deconvolution operation as described in Section \ref{method_decon}. Additionally, as it is already established in the literature, the Delaigle \& Meister estimator \cite{delaigle2008density} for heteroscedastic data was also calculated. The three candidate distributions for $X$ as described in Section \ref{Phase_Func_Sim} were considered. Both normal and Laplace distributions were considered for the measurement error, each in conjunction with the three measurement error variance models outlined in Table \ref{table:error} being considered. In all cases the sample size was taken to be $n=500$. Due to the computational cost of evaluating the phase function deconvolution estimators, a total of $500$ samples were generated for each combination of $X$-distribution and variance model. For the phase-function estimators, the approximate AMISE bandwidth minimizing \eqref{bandwidth_calc} was computed. The bandwidth of the Delaigle-Meister estimator was a two-stage plug-in bandwidth as suggested in their paper. For all the three deconvolution estimators, the integrated squared error (ISE) was computed for each sample. 

Table \ref{table:deconv sim} presents the simulation results corresponding to the setting where the measurement error variances are assumed known, and Table \ref{table:deconv sim_2} presents the simulation results corresponding to the case with $J=2$ replicates per observation and the variance components are estimated as outlined in Section \ref{Estimating Variances}. The simulation with replicate observations contains results for the Delaigle-Meister estimator both using the estimated variances (D\&M$_{\mathrm{VarE}}$) and treating the variances as known (D\&M$_{\mathrm{VarK}}$). Note that the simulations with replicate observations use the individual-level average data $W_i = (W_{i1}+W_{i2})/2$ to compute the deconvolution estimators and are therefore not directly comparable to the simulation without replication and measurement error variances assumed known. Due to the presence of outliers in the ISE calculations, the median as well as the first and third quartiles of $10 \times \mathrm{ISE}$ are reported.

\begin{table}[h]
	\centering
	\begin{tabular}{ccccccHHHH}	
		\hline
		True X  & Error type & Error case &  EPF & WEPF$_{opt}$ & D\&M &&&&\\ 
		\hline
		
		\cline{4-10}	
		Scaled $\chi^2_3$ & Normal &   1 & 0.225 & 0.199 & 0.193 & 0.204 & 0.192 & 0.178 & 0.274 \\ 
		&  &  & [0.189,  0.282] & [0.159,  0.240] & [0.166,  0.230] & [0.164, 0.259] & [0.156, 0.241] & [0.154, 0.205] & [0.233, 0.319] \\ 
		&  &   2 & 0.483 & 0.482 & 0.458 & 0.321 & 0.322 & 0.336 & 0.423 \\ 
		&  &  & [0.404,  0.581] & [0.392,  0.571] & [0.386,  0.547] & [0.252, 0.387] & [0.267, 0.385] & [0.28, 0.405] & [0.384, 0.474] \\ 
		&  &   3 & 0.419 & 0.366 & 0.315 & 0.29 & 0.285 & 0.249 & 0.384 \\ 
		&  &  & [0.321,  0.493] & [0.296,  0.421] & [0.264,  0.39] & [0.234, 0.327] & [0.237, 0.33] & [0.21, 0.298] & [0.335, 0.419] \\ 
		& Laplace &   1 & 0.191 & 0.172 & 0.181 & 0.176 & 0.165 & 0.148 & 0.209 \\ 
		&  &  & [0.167,  0.245] & [0.147,  0.210] & [0.145,  0.213] & [0.142, 0.216] & [0.140, 0.207] & [0.123, 0.180] & [0.176, 0.246] \\ 
		& &   2 & 0.311 & 0.306 & 0.299 & 0.277 & 0.273 & 0.281 & 0.343 \\ 
		&  &  & [0.243,  0.392] & [0.236,  0.371] & [0.229,  0.367] & [0.223, 0.349] & [0.222, 0.337] & [0.234, 0.338] & [0.301, 0.378] \\ 
		& &   3 & 0.27 & 0.268 & 0.266 & 0.219 & 0.218 & 0.23 & 0.298 \\ 
		&  &  & [0.224,  0.352] & [0.205,  0.339] & [0.222,  0.325] & [0.18, 0.266] & [0.176, 0.267] & [0.184, 0.276] & [0.249, 0.325] \\ 
		Mixture 1 & Normal &   1 & 0.184 & 0.140 & 0.117 & 0.128 & 0.120 & 0.097 & 0.206 \\ 
		&  &  & [0.128,  0.248] & [0.085,  0.194] & [0.082,  0.155] & [0.088, 0.182] & [0.077, 0.166] & [0.062, 0.145] & [0.162, 0.277] \\ 
		& &   2 & 0.605 & 0.555 & 0.527 & 0.31 & 0.309 & 0.308 & 0.464 \\ 
		&  &  & [0.452,  0.723] & [0.433,  0.715] & [0.416,  0.63] & [0.214, 0.387] & [0.217, 0.4] & [0.232, 0.401] & [0.404, 0.534] \\ 
		& &   3 & 0.436 & 0.385 & 0.304 & 0.257 & 0.242 & 0.195 & 0.374 \\ 
		&  &  & [0.319,  0.566] & [0.271,  0.503] & [0.182,  0.401] & [0.175, 0.345] & [0.182, 0.339] & [0.12, 0.266] & [0.309, 0.451] \\ 
		& Laplace &   1 & 0.142 & 0.107 & 0.105 & 0.102 & 0.105 & 0.082 & 0.147 \\ 
		&  &  & [0.078,  0.201] & [0.060,  0.160] & [0.073,  0.141] & [0.066, 0.156] & [0.074, 0.159] & [0.058, 0.117] & [0.106, 0.199] \\ 
		& &   2 & 0.265 & 0.258 & 0.242 & 0.216 & 0.21 & 0.223 & 0.308 \\ 
		&  &  & [0.19,  0.384] & [0.182,  0.354] & [0.156,  0.326] & [0.151, 0.271] & [0.14, 0.267] & [0.154, 0.272] & [0.255, 0.355] \\ 
		& &   3 & 0.254 & 0.232 & 0.212 & 0.193 & 0.176 & 0.161 & 0.267 \\ 
		&  &  & [0.178,  0.339] & [0.173,  0.293] & [0.142,  0.271] & [0.13, 0.283] & [0.119, 0.242] & [0.114, 0.244] & [0.229, 0.333] \\ 
		Mixture 2 & Normal &   1 & 0.098 & 0.090 & 0.073 & 0.081 & 0.084 & 0.064 & 0.123 \\ 
		&  &  & [0.063,  0.175] & [0.051,  0.136] & [0.053,  0.105] & [0.055, 0.111] & [0.051, 0.110] & [0.049, 0.088] & [0.098, 0.150] \\ 
		&&   2 & 0.296 & 0.296 & 0.274 & 0.189 & 0.185 & 0.164 & 0.247 \\ 
		&  &  & [0.224,  0.387] & [0.21,  0.391] & [0.201,  0.343] & [0.112, 0.251] & [0.118, 0.243] & [0.126, 0.227] & [0.204, 0.285] \\ 
		& &   3 & 0.223 & 0.2 & 0.172 & 0.132 & 0.125 & 0.118 & 0.201 \\ 
		&  &  & [0.152,  0.286] & [0.132,  0.26] & [0.118,  0.217] & [0.096, 0.193] & [0.082, 0.194] & [0.077, 0.144] & [0.172, 0.239] \\ 
		& Laplace &   1 & 0.073 & 0.073 & 0.070 & 0.070 & 0.070 & 0.056 & 0.087 \\ 
		&  &  & [0.049,  0.128] & [0.044,  0.107] & [0.041,  0.104] & [0.049, 0.101] & [0.046, 0.099] & [0.037, 0.082] & [0.059, 0.122] \\ 
		& &   2 & 0.154 & 0.146 & 0.164 & 0.136 & 0.117 & 0.15 & 0.181 \\ 
		&  &  & [0.1,  0.22] & [0.1,  0.23] & [0.103,  0.239] & [0.086, 0.187] & [0.077, 0.163] & [0.106, 0.186] & [0.156, 0.214] \\ 
		& &   3 & 0.139 & 0.125 & 0.141 & 0.117 & 0.103 & 0.125 & 0.169 \\ 
		&  &  & [0.096,  0.189] & [0.081,  0.174] & [0.101,  0.192] & [0.076, 0.175] & [0.073, 0.165] & [0.086, 0.168] & [0.138, 0.208] \\ 
		\hline
	\end{tabular}
	
	\caption{Density estimation for $n=500$ with no replicates and measurement error variances are assumed to be known. The median, as well as first and third quartiles, [$Q_1, Q_3$], of 10 $ \times$ ISE of density estimators under 500 simulations.} 
	\label{table:deconv sim}
	
\end{table}
\newcolumntype{x}[1]{>{\centering\arraybackslash\hspace{0pt}}p{#1}}
\begin{table}[h]
	\centering
	\begin{tabular}{cccHHHx{2.2cm}x{2.2cm}x{2.2cm}x{2.2cm}}	
		\hline
		True X  & Error type & Error case & & & &  EPF & WEPF$_{opt}$ & D\&M$_\mathrm{VarK}$  & D\&M$_\mathrm{VarE}$  \\ 
		\hline
		
		\cline{4-10}	
		Scaled $\chi^2_3$ & Normal &   1 & 0.225 & 0.199 & 0.193 & 0.204 & 0.192 & 0.178 & 0.274 \\ 
		&  &  & [0.189,  0.282] & [0.159,  0.240] & [0.166,  0.230] & [0.164, 0.259] & [0.156, 0.241] & [0.154, 0.205] & [0.233, 0.319] \\ 
		&  &   2 & 0.483 & 0.482 & 0.458 & 0.321 & 0.322 & 0.336 & 0.423 \\ 
		&  &  & [0.404,  0.581] & [0.392,  0.571] & [0.386,  0.547] & [0.252, 0.387] & [0.267, 0.385] & [0.28, 0.405] & [0.384, 0.474] \\ 
		&  &   3 & 0.419 & 0.366 & 0.315 & 0.29 & 0.285 & 0.249 & 0.384 \\ 
		&  &  & [0.321,  0.493] & [0.296,  0.421] & [0.264,  0.39] & [0.234, 0.327] & [0.237, 0.33] & [0.21, 0.298] & [0.335, 0.419] \\ 
		& Laplace &   1 & 0.191 & 0.172 & 0.181 & 0.176 & 0.165 & 0.148 & 0.209 \\ 
		&  &  & [0.167,  0.245] & [0.147,  0.210] & [0.145,  0.213] & [0.142, 0.216] & [0.140, 0.207] & [0.123, 0.180] & [0.176, 0.246] \\ 
		& &   2 & 0.311 & 0.306 & 0.299 & 0.277 & 0.273 & 0.281 & 0.343 \\ 
		&  &  & [0.243,  0.392] & [0.236,  0.371] & [0.229,  0.367] & [0.223, 0.349] & [0.222, 0.337] & [0.234, 0.338] & [0.301, 0.378] \\ 
		& &   3 & 0.27 & 0.268 & 0.266 & 0.219 & 0.218 & 0.23 & 0.298 \\ 
		&  &  & [0.224,  0.352] & [0.205,  0.339] & [0.222,  0.325] & [0.18, 0.266] & [0.176, 0.267] & [0.184, 0.276] & [0.249, 0.325] \\ 
		Mixture 1 & Normal &   1 & 0.184 & 0.140 & 0.117 & 0.128 & 0.120 & 0.097 & 0.206 \\ 
		&  &  & [0.128,  0.248] & [0.085,  0.194] & [0.082,  0.155] & [0.088, 0.182] & [0.077, 0.166] & [0.062, 0.145] & [0.162, 0.277] \\ 
		& &   2 & 0.605 & 0.555 & 0.527 & 0.31 & 0.309 & 0.308 & 0.464 \\ 
		&  &  & [0.452,  0.723] & [0.433,  0.715] & [0.416,  0.63] & [0.214, 0.387] & [0.217, 0.4] & [0.232, 0.401] & [0.404, 0.534] \\ 
		& &   3 & 0.436 & 0.385 & 0.304 & 0.257 & 0.242 & 0.195 & 0.374 \\ 
		&  &  & [0.319,  0.566] & [0.271,  0.503] & [0.182,  0.401] & [0.175, 0.345] & [0.182, 0.339] & [0.12, 0.266] & [0.309, 0.451] \\ 
		& Laplace &   1 & 0.142 & 0.107 & 0.105 & 0.102 & 0.105 & 0.082 & 0.147 \\ 
		&  &  & [0.078,  0.201] & [0.060,  0.160] & [0.073,  0.141] & [0.066, 0.156] & [0.074, 0.159] & [0.058, 0.117] & [0.106, 0.199] \\ 
		& &   2 & 0.265 & 0.258 & 0.242 & 0.216 & 0.21 & 0.223 & 0.308 \\ 
		&  &  & [0.19,  0.384] & [0.182,  0.354] & [0.156,  0.326] & [0.151, 0.271] & [0.14, 0.267] & [0.154, 0.272] & [0.255, 0.355] \\ 
		& &   3 & 0.254 & 0.232 & 0.212 & 0.193 & 0.176 & 0.161 & 0.267 \\ 
		&  &  & [0.178,  0.339] & [0.173,  0.293] & [0.142,  0.271] & [0.13, 0.283] & [0.119, 0.242] & [0.114, 0.244] & [0.229, 0.333] \\ 
		Mixture 2 & Normal &   1 & 0.098 & 0.090 & 0.073 & 0.081 & 0.084 & 0.064 & 0.123 \\ 
		&  &  & [0.063,  0.175] & [0.051,  0.136] & [0.053,  0.105] & [0.055, 0.111] & [0.051, 0.110] & [0.049, 0.088] & [0.098, 0.150] \\ 
		&&   2 & 0.296 & 0.296 & 0.274 & 0.189 & 0.185 & 0.164 & 0.247 \\ 
		&  &  & [0.224,  0.387] & [0.21,  0.391] & [0.201,  0.343] & [0.112, 0.251] & [0.118, 0.243] & [0.126, 0.227] & [0.204, 0.285] \\ 
		& &   3 & 0.223 & 0.2 & 0.172 & 0.132 & 0.125 & 0.118 & 0.201 \\ 
		&  &  & [0.152,  0.286] & [0.132,  0.26] & [0.118,  0.217] & [0.096, 0.193] & [0.082, 0.194] & [0.077, 0.144] & [0.172, 0.239] \\ 
		& Laplace &   1 & 0.073 & 0.073 & 0.070 & 0.070 & 0.070 & 0.056 & 0.087 \\ 
		&  &  & [0.049,  0.128] & [0.044,  0.107] & [0.041,  0.104] & [0.049, 0.101] & [0.046, 0.099] & [0.037, 0.082] & [0.059, 0.122] \\ 
		& &   2 & 0.154 & 0.146 & 0.164 & 0.136 & 0.117 & 0.15 & 0.181 \\ 
		&  &  & [0.1,  0.22] & [0.1,  0.23] & [0.103,  0.239] & [0.086, 0.187] & [0.077, 0.163] & [0.106, 0.186] & [0.156, 0.214] \\ 
		& &   3 & 0.139 & 0.125 & 0.141 & 0.117 & 0.103 & 0.125 & 0.169 \\ 
		&  &  & [0.096,  0.189] & [0.081,  0.174] & [0.101,  0.192] & [0.076, 0.175] & [0.073, 0.165] & [0.086, 0.168] & [0.138, 0.208] \\ 
		\hline
	\end{tabular}
	
	\caption{Density estimation for $n=500$ with $J=2$ replicates for each observation. The median, as well as first and third quartiles, [$Q_1, Q_3$], of 10 $ \times$ ISE of density estimators under 500 simulations.} 
	\label{table:deconv sim_2}
	
\end{table}

Inspection of Table \ref{table:deconv sim} reveals that the Delaigle-Meister (D\&M) estimator tends to have the smallest median ISE, although there are a few instances in which the phase function estimators outperform the D\&M estimator, notably for Mixture 2 and Laplace measurement error. It is also clear that calculating the mean-optimal weights is very advantageous in this setting, with the mean-optimally weighted estimator having smaller median ISE than the equally weighted estimator in all but one instance. Overall, one can conclude that the WEPF estimator performs very well and compares favorably to the D\&M estimator, the latter requiring knowledge of the measurement error distribution to be useful in practice.

Inspection of the simulation results in Table \ref{table:deconv sim_2} is very insightful. Note that the measurement error variances here are estimated based on only $J=2$ replicates for each observation. As such, one might not expect good performance. However, the two phase function estimators perform very favorable when compared to the D\&M estimator with known measurement error variances. The mean-optimally weighted estimator generally performs better than the equally weighted estimators in terms of median ISE, although there are two exceptions. It is interesting that weights estimated based on only two replicates give such good performance. Also revealing is that the WEPF estimator performs significantly better than the D\&M estimator with estimated variances, with the median ISE of the mean-optimally weighted estimator often reflecting more than a $50\%$ reduction in median ISE when comapared to the D\&M counterpart.

\begin{landscape}
	\begin{figure}
		\begin{tabular}{ccc}
			\includegraphics[width=70mm,page=1]{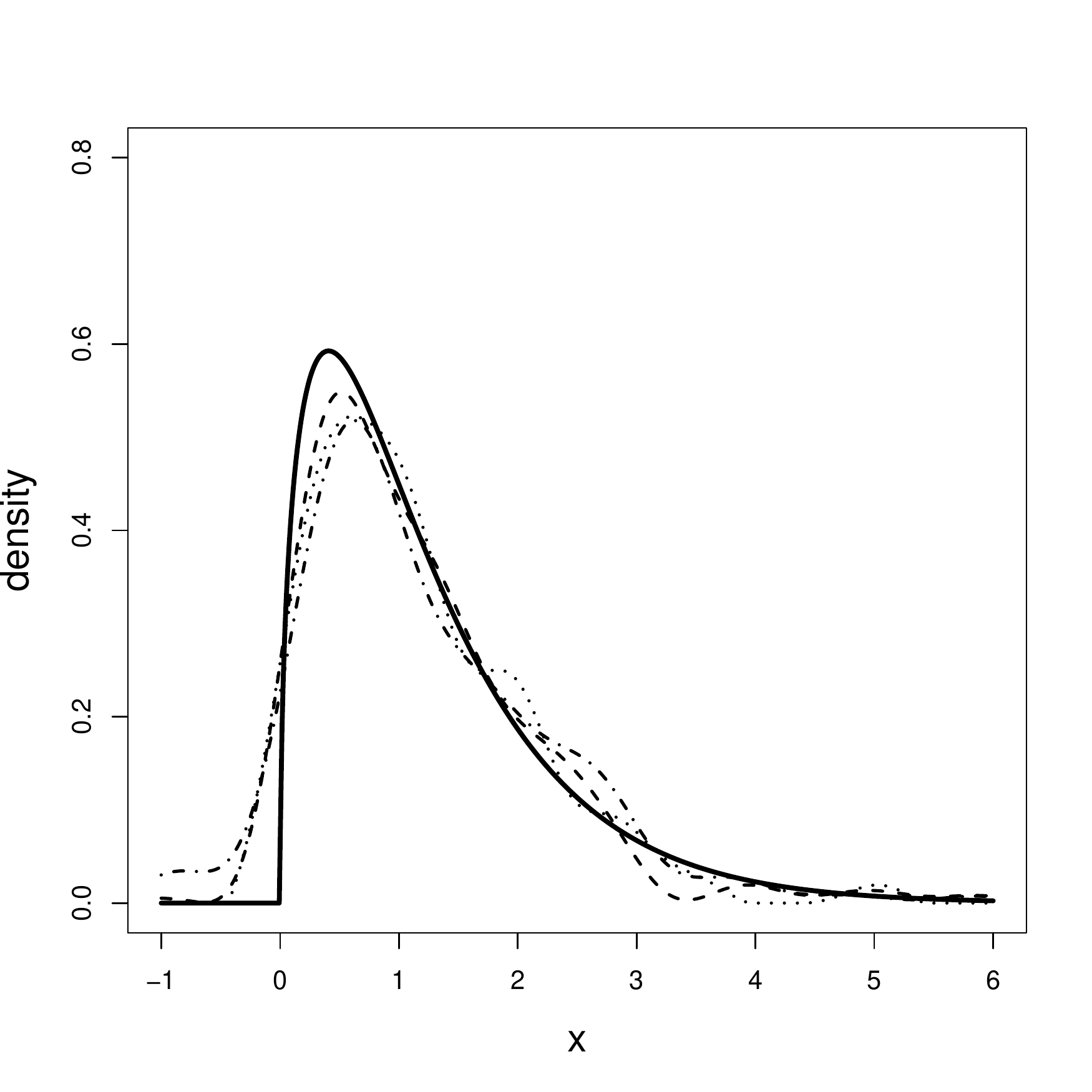} & \includegraphics[width=70mm,page=2]{density_plot.pdf} & 
			\includegraphics[width=70mm,page=4]{density_plot.pdf} \\
			(a) & (b) & (c) \\
			\includegraphics[width=70mm,page=9]{density_plot.pdf} & \includegraphics[width=70mm,page=10]{density_plot.pdf} & 
			\includegraphics[width=70mm,page=12]{density_plot.pdf} \\
			(d) & (e) & (f)
		\end{tabular}
		\caption{Curves Q$_1$ (\protect\tikz[baseline]{\protect\draw[line width=0.1mm,dashed] (0,.8ex)--++(1,0);}), Q$_2$ (\protect\tikz[baseline]{\protect\draw[line width=0.1mm,dotted] (0,.8ex)--++(1,0);}), Q$_3$ (\protect\tikz[baseline]{\protect\draw[line width=0.1mm,dash dot] (0,.8ex)--++(1,0);}), and true curve (\protect\tikz[baseline]{\protect\draw[line width=0.1mm] (0,.8ex)--++(1,0);})  for $X \sim $ Scaled-$\chi^2_3$, $n=500$, $J=2$ replicates per observation when the errors are Normal (a)-(c), and Laplace (d)-(f), with case 1 of measurement error variances. For (a),(d): EPF estimator; (b),(e): WEPF$_{\textrm{opt}}$ estimator; (c),(f): D\&M estimator with estimated variances. All estimators are computed using plug-in bandwidth. }	
		\label{fig:den-chisq}
	\end{figure}
\end{landscape}

\begin{landscape}
	\begin{figure}
		\begin{tabular}{ccc}
			\includegraphics[width=70mm,page=17]{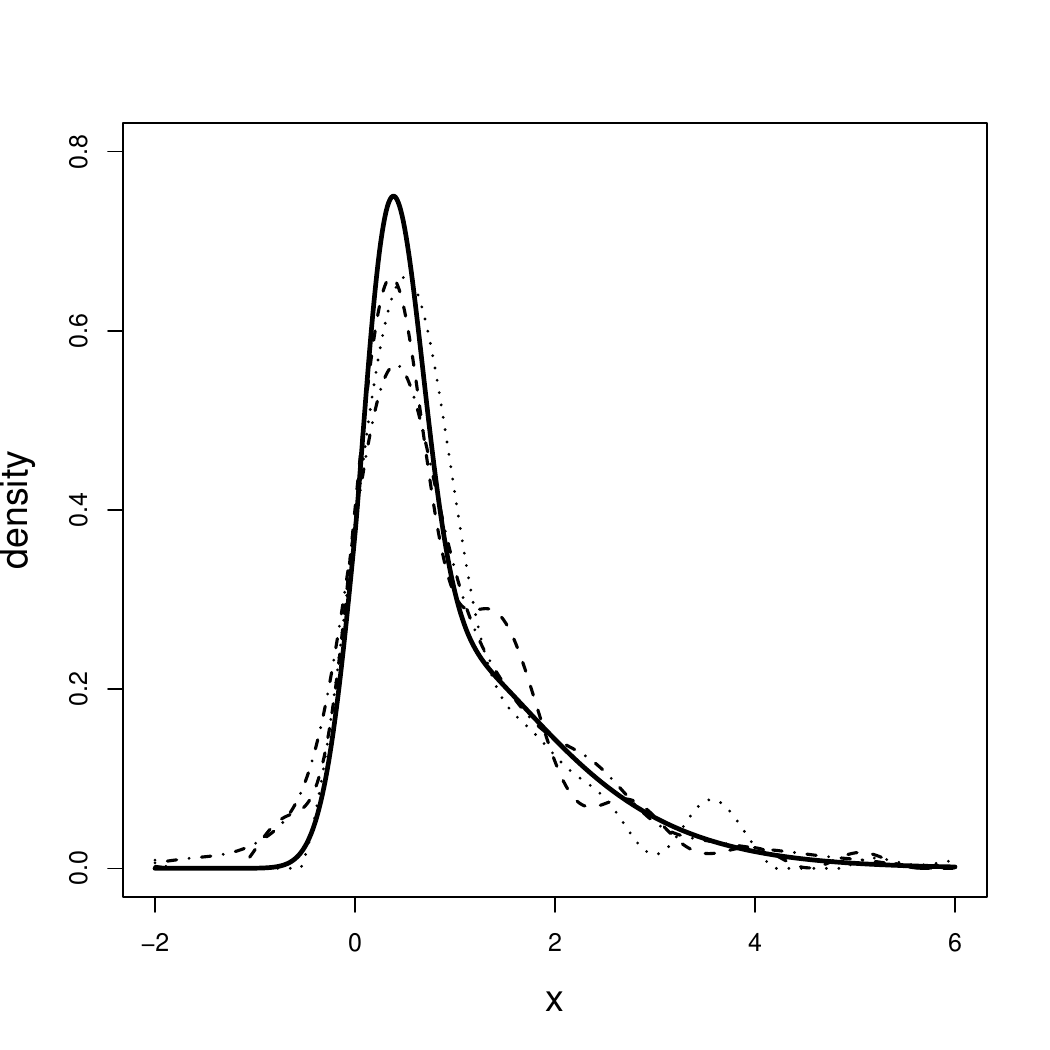} & \includegraphics[width=70mm,page=18]{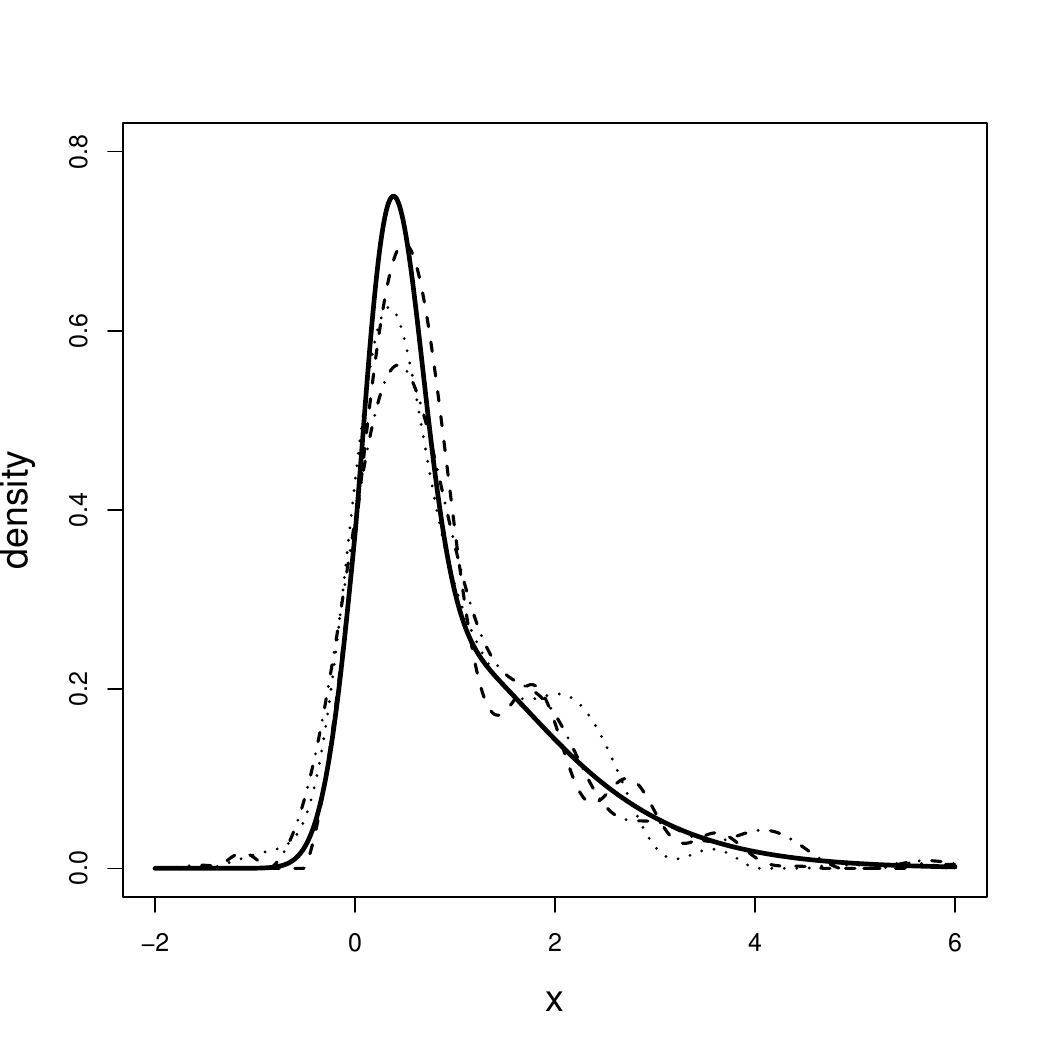} & 
			\includegraphics[width=70mm,page=20]{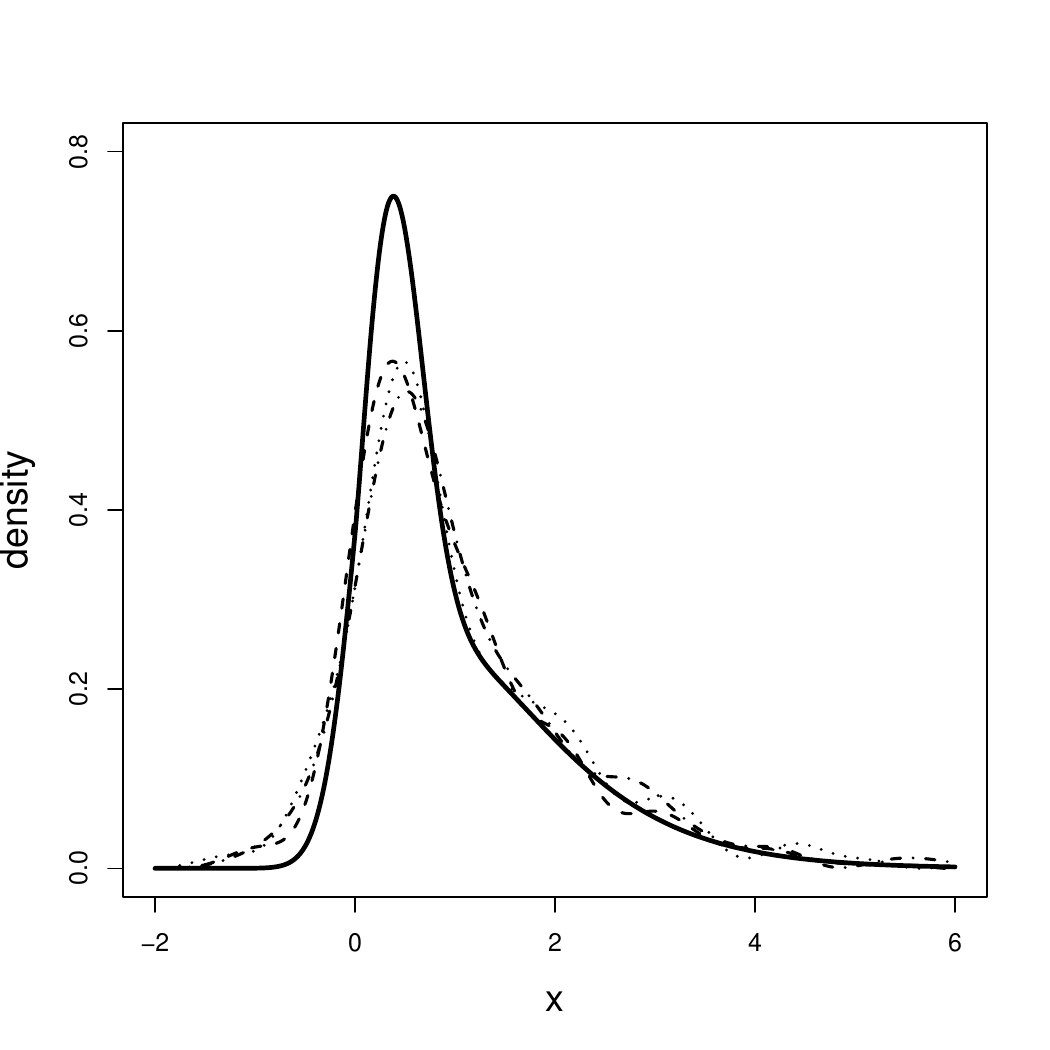} \\
			(a) & (b) & (c) \\
			
			\includegraphics[width=70mm,page=25]{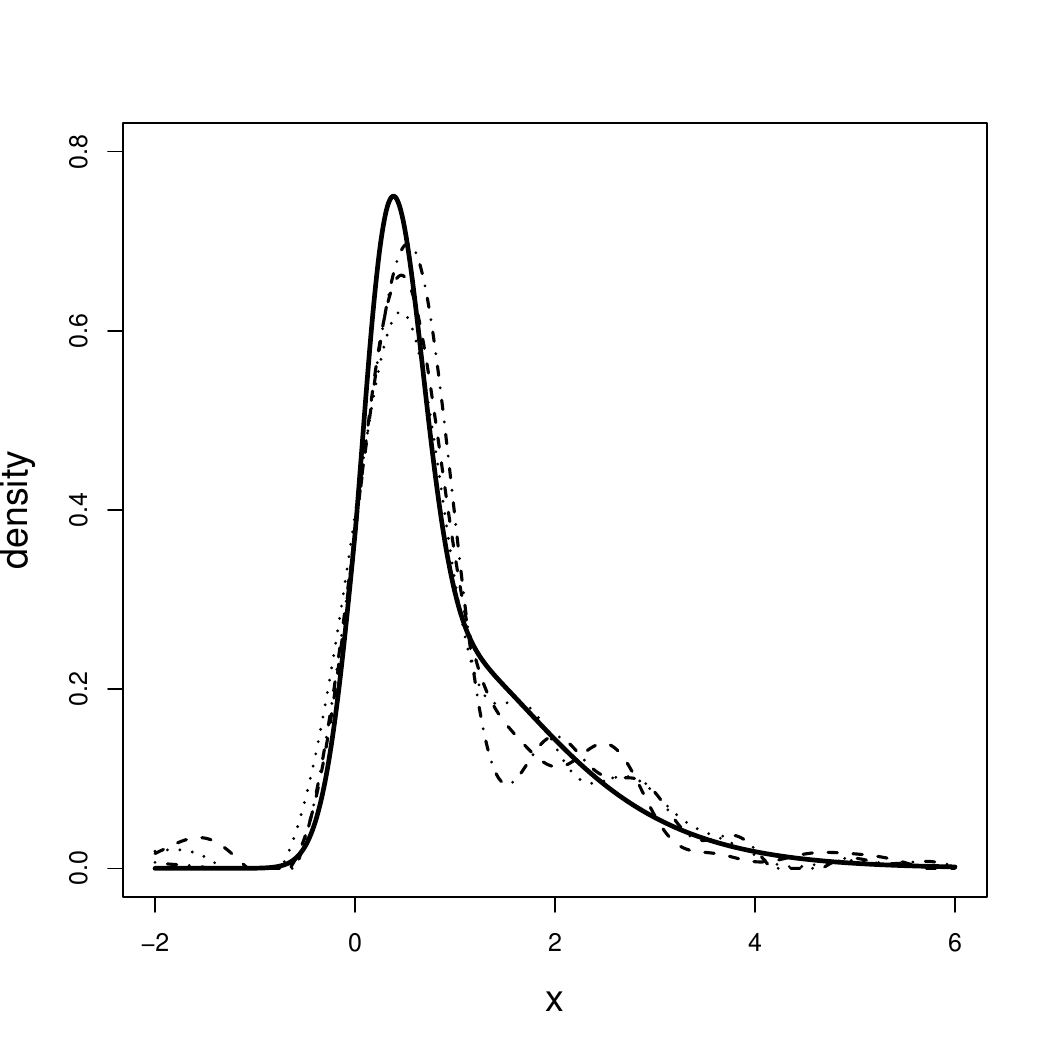} & \includegraphics[width=70mm,page=26]{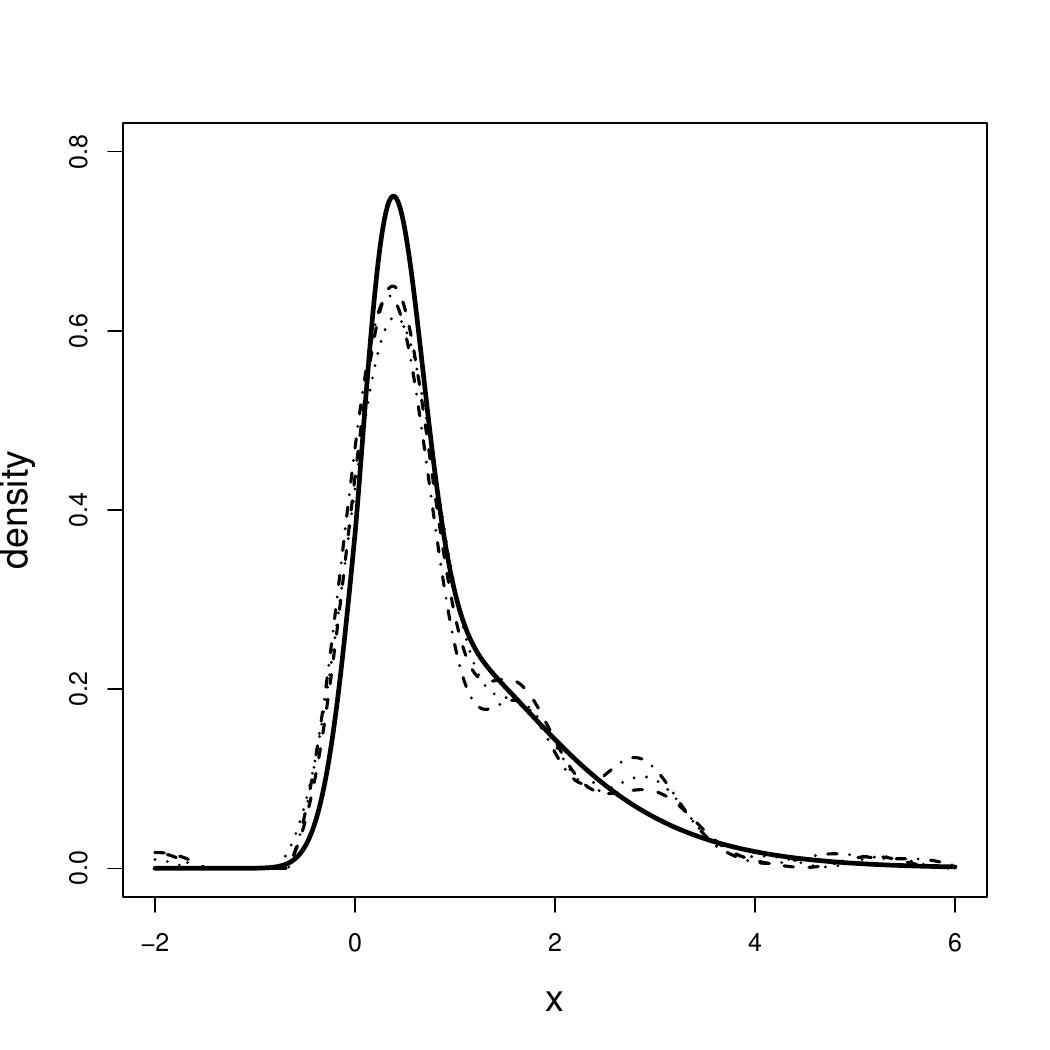} & 
			\includegraphics[width=70mm,page=28]{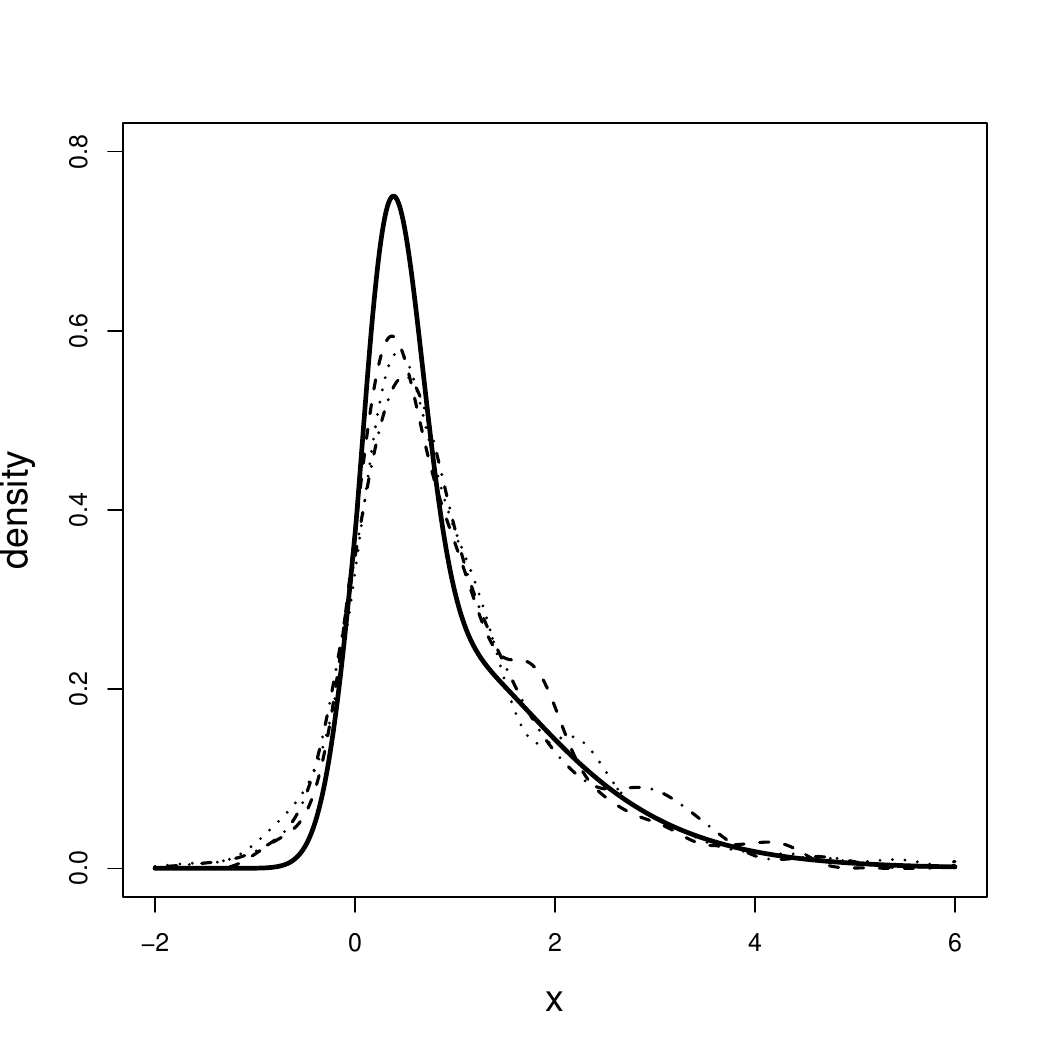} \\
			(d) & (e) & (f)
		\end{tabular}
		\caption{Curves Q$_1$ (\protect\tikz[baseline]{\protect\draw[line width=0.1mm,dashed] (0,.8ex)--++(1,0);}), Q$_2$ (\protect\tikz[baseline]{\protect\draw[line width=0.1mm,dotted] (0,.8ex)--++(1,0);}), Q$_3$ (\protect\tikz[baseline]{\protect\draw[line width=0.1mm,dash dot] (0,.8ex)--++(1,0);}), and true curve (\protect\tikz[baseline]{\protect\draw[line width=0.1mm] (0,.8ex)--++(1,0);})  for $X \sim $ Mixture 1, $n=500$, $J=2$ replicates per observation, when the errors are Normal (a)-(c), and Laplace (d)-(f), with case 1 of measurement error variances. For (a),(d): EPF estimator; (b),(e): WEPF$_{\textrm{opt}}$ estimator; (c),(f): D\&M estimator with estimated variances. All estimators are computed using plug-in bandwidth. }
		\label{fig:den-mix1}	
	\end{figure}
\end{landscape}

Figures \ref{fig:den-chisq} and \ref{fig:den-mix1} show plots of the density estimators corresponding to the first, second, and third quantiles (Q$_1$, Q$_2$, and Q$_3$) of ISE for each of the methods EPF, WEPF$_{opt}$, and the D\&M estimators corresponding to $X$ having scaled $\chi^2_3$ and Mixture 1 distribution. In all three instances, the estimators were calculated with estimated measurement error variances based on $J=2$ replicates per observation. Observation-level measurement error was taken to be Case 1 of Table \ref{table:error}. Both normal and Laplace distributions were considered for the measurement error. The sample size was fixed at $n=500$. The figures also show the true density curve for comparison. Although all three estimators considered are able to capture the shape of the true density, the D\&M estimators with estimated variance do the worst among the three: For $X$ having a scaled $\chi^2_3$ distribution, it puts much more density in negative support than the EPF and WEPF$_{\mathrm{opt}}$ and tends to underestimate the modal height. Both the EPF with WEPF$_{\mathrm{opt}}$, perform well for the scaled $\chi_3^2$ distribution, with the WEPF$_{opt}$ seemingly capturing the shape around the mode a little better than the EPF. When evaluating Figure \ref{fig:den-mix1} showing the same plots for $X$ having the distribution Mixture 1, the general observations are very similar. The EPF and WEPF$_{opt}$ have visually similar performance, while the D\&M estimator underestimates the density around the mode. The Supplementary Material also contains a set of plots corresponding to $X$ having Mixture 2 distribution.  Similar observations apply there.

Additional simulation results are presented in the Supplemental Material. There, the  EPF, WEPF and D\&M estimators are compared under the assumption that one can find an optimal bandwidth (a bandwidth minimizing ISE) for any observed sample. When no replicate data is available and the measurement error variances are assumed known, the D\&M estimator has the best performance, and the WEPF outperforms the EPF in all but one case considered. However, once the measurement error variance needs to be estimated (for both $J=2$ and $J=3$ replicates per case), the WEPF estimator tends to have the best performance, with the D\&M estimator faring worse than the EPF estimator. Finally, a simulation with plug-in bandwidth and $J=3$ replicates is also presented. Here, the EPF and WEPF both outperform the D\&M estimator.

\section{Analysis of Framingham Data}

In this section, the EPF and WEPF$_{opt}$ density deconvolution estimators are illustrated using a classical dataset in the deconvolution literature, a subset of the Framingham Heart Study. The data consists of several variables related to coronary heart disease for $n=1615$ patients. For each patient, two measurements of long-term systolic blood pressure (SBP) were collected at each of two examination. As per Carroll et al., \cite{carroll2006measurement} let $M_{ij}$ be the average of the two measurements at exam $j$ for $j=1,2$, and let $W_{ij} = \log(M_{ij}-50)$. The $W_{ij}$ are assumed to be related to true long-term SBP, $X_i$ according to $W_{ij} = Y_i + \sigma_i \varepsilon_{ij}$ with $Y_i = \log(X_i-50)$. Density deconvolution is therefore used to estimate the density on the $Y$-scale, $\hat{f}_Y(y)$, after which it follows that $\hat{f}_X(x)=(x-50)^{-1}\hat{f}_Y[\log(x-50)]$, $x>50$.

\begin{figure}[h]
	\centering
	\includegraphics[scale=0.75]{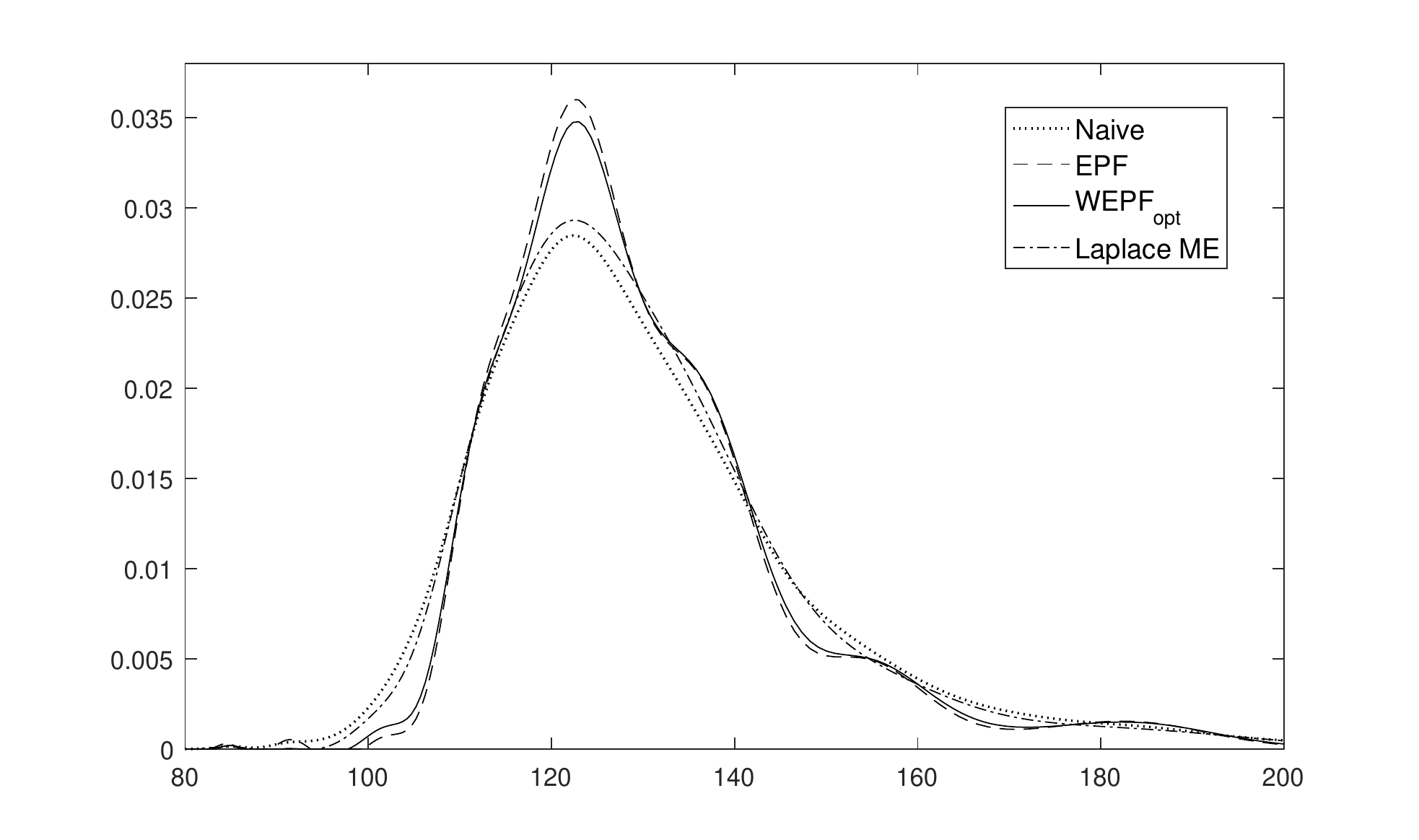}
	\caption{Estimation of the density $f_X$ in the Framingham data. Four density estimates are shown: a naive kernel estimator (measurement error is ignored), the EPF estimator, the WEPF$_{opt}$ estimator, and the Delaigle \& Meister estimator assuming Laplace measurement error.}
	\label{fig:Fram}
\end{figure}

For the SBP data, the EPF and WEPF$_{opt}$ were estimated, the latter with mean-optimal weights $\boldsymbol{q}_{opt}$ using variance components estimated as described in Section \ref{Estimating Variances}. For both the EPF and WEPF$_{opt}$, deconvolution bandwidths were estimated using \eqref{bandwidth_calc}. These two estimators are shown in Figure \ref{fig:Fram}, together with the Delaigle \& Meister (2008) estimator using the same estimated variances and Laplace measurement error. (The D\&M estimator was also calculated for normal measurement error and was nearly identical.) A naive kernel estimator of the data using a normal references bandwidth is also shown for comparative purposes.Other bandwidth selection approaches for the naive kernel estimator were also considered with very similar results. The naive kernel estimator is much flatter around the mode and fatter in the tails. This is expected, as the kernel estimator makes no correction for the measurement error present in the data. Furthermore, it can be seen that the WEPF$_{opt}$ and EPF deconvolution density estimators are similar. The two density estimators based on phase functions suggest that the distribution of $X$ may be multi-modal, while the D\&M estimator is unimodal and positive skew.

\section{Conclusions}
This paper presents a method for phase density deconvolution with heteroscedastic measurement error of unknown type and builds on the work of Delaigle \& Hall \cite{delaigle2016methodology} who considered the homoscedastic case. Two estimators are proposed, one using equally weighted observations and the other using mean-optimal weights to adjust for heteroscedasticity of the measurement error. A method based on approximating the AMISE is proposed for bandwidth selection in both instances. In the simulation settings considered, the WEPF$_{opt}$ estimator generally performed better than the EPF estimator, although there were instances where their performance was comparable. The simulation results suggest that mean-optimal weighting of observations will not have a detrimental effect on estimating the density function, and big gains are sometimes possible. The practitioner cautious about estimaging weights from a small number of replicates could always opt for a hybrid type of estimator, calculting WEPF$_{hybrid}$ using weights $\boldsymbol{q}_{\mathrm{hybrid}} = \alpha \boldsymbol{q}_{\mathrm{opt}} + (1-\alpha)/n$ where $\alpha$ indicates their degree of confidence in using the estimated weights. The performance of this hybrid estimator is a future avenue of research. In the setting where the measurement error variances are known, the method of Delaigle \& Meister \cite{delaigle2008density} will outperform both phase function estimators, although the latter are still competitive in this setting. Also recall that the Delaigle \& Meister estimator requires knowledge of the measurement error distribution --- an assumption not made by the EPF and WEPF estimators. When there are only 2 replicates per individual from which to estimate the measurement error variances, the phase function methods performed substantially better than the Delaigle \& Meister estimator. This suggests that the phase function methods have some inherent robustness against variance estimate deviation from the true values, and that the phase function density estimators can generally do the same as Delaigle \& Meister estimator with much less assumption on measurement error.  

\section*{Supplementary Material}

In the supplementary material, the asymptotic properties of the weighted empirical phase function (WEPF) and the mean integrated squared error (MISE) of the phase function deconvolution density estimator are derived. Furthermore, plots of the phase functions corresponding to the three distributions used in the simulation studies (Section \ref{Phase_Func_Sim}) are shown. In addition, as a complement to the simulations in Section \ref{Sim_Dens}, the plots of the density estimators corresponding to the first, second, and third quantiles (Q$_1$, Q$_2$, and Q$_3$) of ISE for each of the methods EPF, WEPF$_{opt}$, and the D\&M estimators corresponding to $X$ having a bimodal mixture distribution (called Mixture 2 in the paper).	Finally, simulation results are provided to compare density estimators under an optimal bandwidth setting and also when there are $J=3$ replicates per observation.


\end{document}